\documentclass[10pt,journal,compsoc]{IEEEtran}
\usepackage{amsmath,amsfonts}
\usepackage{algorithmic}
\usepackage{algorithm}
\usepackage{array}
\usepackage[caption=false,font=normalsize,labelfont=sf,textfont=sf]{subfig}
\usepackage{textcomp}
\usepackage{stfloats}
\usepackage{url}
\usepackage{verbatim}
\usepackage{graphicx}
\usepackage{cite}
\usepackage{xspace}
\usepackage{xcolor}
\usepackage{hyperref}
\newcommand{\myparagraph}[1]{\vspace{1mm} \noindent \textbf{#1}}
\newcommand{\ie}{i.e.,\xspace}
\newcommand{\etal}{et al.\xspace}
\newcommand{\newText}[1]{\textcolor{black}{#1}\xspace}
\newcommand{\newtext}[1]{\textcolor{black}{#1}\xspace}
\newcommand{\newtextF}[1]{\textcolor{black}{#1}\xspace}

\begin{document}

\title{Continuous Scatterplot and Image Moments for Time-Varying Bivariate Field Analysis of Electronic Structure Evolution}

\author{Mohit Sharma, Talha Bin Masood, Nanna Holmgaard List, Ingrid Hotz and\\ Vijay Natarajan,~\IEEEmembership{Member,~IEEE}
\IEEEcompsocitemizethanks{\IEEEcompsocthanksitem M. Sharma, T.B. Masood and I. Hotz are with the Scientific Visualization group, Department of Science and Technology (ITN), Link\"oping University, Norrk\"oping, Sweden.\protect\\
E-mail: \{mohit.sharma, talha.bin.masood, ingrid.hotz\}@liu.se.
\IEEEcompsocthanksitem  N.H. List is with the Department of Chemistry, KTH Royal Institute of Technology, Stockholm, Sweden.\protect\\
E-mail: nalist@kth.se.
\IEEEcompsocthanksitem  V. Natarajan is with the Department of Computer Science and Automation, Indian Institute of Science, Bangalore.\protect\\
E-mail: vijayn@iisc.ac.in.
}
\thanks{}
}

\markboth{IEEE Transactions on Visualization and Computer Graphics,~Vol.~X, No.~X, Month~Year}%
{Sharma \MakeLowercase{\textit{et al.}}: Continuous Scatterplot and Image Moments for Time-Varying Bivariate Field Analysis of Electronic Structure Evolution}


\IEEEtitleabstractindextext{%
\begin{abstract}
Photoinduced electronic transitions are complex quantum-mechanical processes where electrons move between energy levels due to the absorption of light. This induces dynamics \ie coupled changes in the electronic structure and nuclear geometry, that drive physical and chemical processes of importance in diverse fields ranging from photobiology and materials design to medicine. The evolving electronic structure can be characterized by two electron density fields: hole and particle natural transition orbitals (NTOs). A study of the two density fields helps understand the movement of electronic charge from one part of the molecule to another, specifically the donor and acceptor regions. Previous works in this area rely on side-by-side visual comparisons of isosurfaces, statistical approaches, or visual analysis of bivariate fields restricted to limited time instances. We propose a new method to analyze time-varying bivariate fields with a large number of instances, as pertinent to understand electronic structure changes during light-induced dynamics. Since the NTO fields depend on the nuclear geometry, the nuclear motion leads to a large number of bivariate field instances. Structures like tracking graphs have been used to analyze time-varying univariate fields. This paper presents a structured and practical approach to feature-directed visual exploration of time-varying bivariate fields using continuous scatterplots (CSPs) and image moment-based descriptors, tailored for studying the evolving electronic structure following photoexcitation. The CSP of the bivariate field at every time step is represented using an image moment vector of length $4$. The collection of all image moment vector descriptors is considered as a point cloud in $\mathbb{R}^4$ and visualized using principal component analysis. Choosing an appropriate pair of principal components results in a representation of the point cloud as a curve on the plane. This representation supports tasks such as identifying interesting time steps, identifying patterns within the bivariate field, and tracking their evolution over time. We present two case studies on excited-state dynamics in molecular systems that demonstrate how the time-varying bivariate field analysis helps provide application-specific insights.
\end{abstract}

}

\maketitle

\IEEEdisplaynontitleabstractindextext

\IEEEpeerreviewmaketitle

\begin{figure*}[!t]
\centering
  \includegraphics[width=\linewidth]{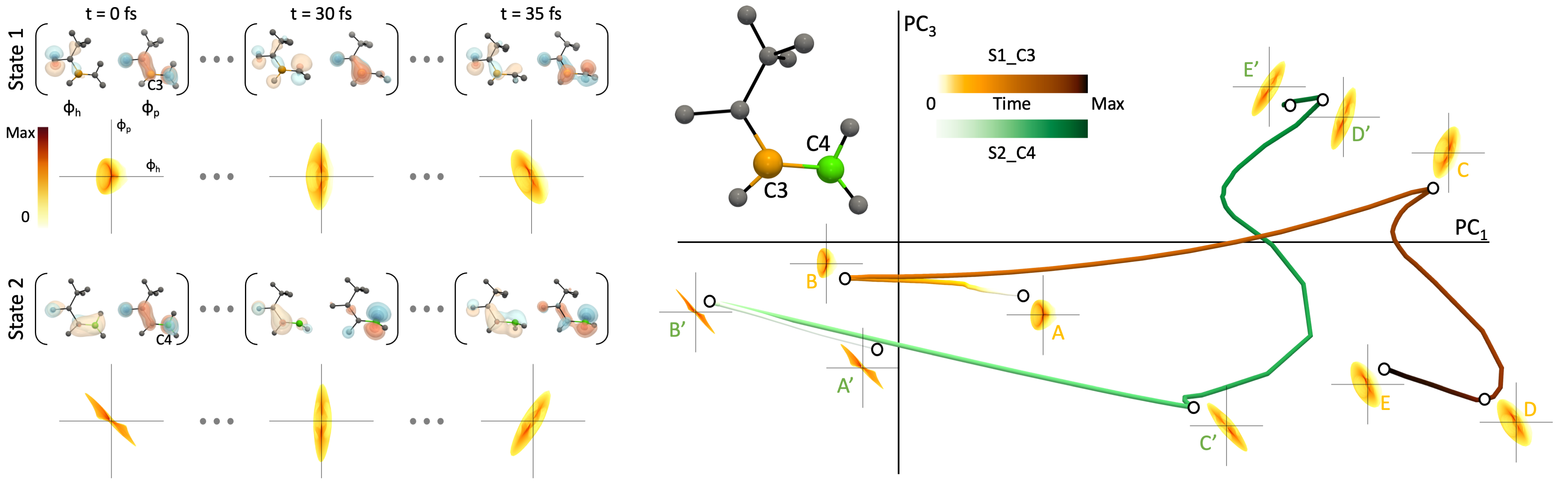}\vspace{-0.4em}
  \caption{Analyzing behavior of two carbon atoms across two electronic transition states. (Left)~Individual electronic density fields ($\phi_h$, $\phi_p$) for both states and CSPs for C3~(orange) and C4~(green) for three time steps. Both CSPs align vertically at $t=30$~fs. (Right)~Molecular structure highlighting the two carbons under consideration, C3  and C4, and tracks in the PC$_1$-PC$_3$ plane for State~1 C3 (S1\_C3, orange) and State~2 C4 (S2\_C4 green). The tracks are annotated at selected time steps with corresponding CSPs. The tracks appear to be similar in the section from $A$ to $E$ and $A'$ to $E'$, with a distinct pattern from $C$ to $D$ and $C'$ to $D'$. Points located nearby have similar CSPs, for example, ($C, D', E'$) or ($C', D, E$).
  }
  \label{fig:teaser}
\end{figure*}

\IEEEraisesectionheading{\section{Introduction}}
\IEEEPARstart{T}{he} study of photoinduced dynamics in molecules is critical to unravel the mechanisms underlying a variety of light-driven processes in both natural and engineered systems \newtextF{(e.g., light-harvesting by plants or charge-separation in photovoltaic materials)}. When a molecule absorbs light, electrons move from occupied orbitals to unoccupied orbitals, resulting in a change in the electronic structure of the molecule. The orbital left vacant by the electron is called the \emph{hole}, while the orbital filled by the electron is called the \emph{particle}. The movement of charge depends on the geometry of the molecule. \newtextF{We use the Natural Transition Orbital~(NTO) decomposition~\cite{martin_nto_2023}, the hole NTO ($\phi_h$) and particle NTO ($\phi_p$), for characterizing changes in electronic structure for a specific geometry. The two NTO fields capture the origin and destination of electrons for a given electronic excited state relative to the ground state. We consider the NTO fields as a time-varying bivariate field across evolving nuclear geometries and develop a method that helps chemists analyze the behavior of the field within volumetric segments such as atoms. Analyzing charge transfer and identifying donor/acceptor regions are important steps in the study of electronic structure changes~\cite{Sharma2023}. In a time-varying scenario, we address the challenge of an evolving molecular geometry by introducing a summary representation that can support analysis tasks such as (a)~Identifying interesting atoms or time steps, (b)~Identifying outliers, (c)~Capturing the evolution of bivariate field within individual atoms, (d)~Identifying atoms with similar behavior, and (e)~Identifying atoms that exhibit similar pattern of evolution. These tasks help the chemist in follow-up studies of physical phenomena related to the identified atoms, time steps, or patterns, which can ultimately aid in the design of novel materials}.

\newtextF{Existing techniques for visualizing donor/acceptor regions typically rely on side-by-side comparisons of isosurfaces overlaid on the ball-and-stick model~\cite{Humphrey1996,Stone2011,Haranczyk2008}, point-wise differences in electron density fields~\cite{Garcia2010,Bahers2011,Guido2013}, computation of aggregate physical quantities or topological descriptors~\cite{masood2021visual,Thygesen2022,Wetzels2024}, or bivariate analysis~\cite{sharma2021segmentation,Sharma2023}. These methods often require manual intervention, and a direct extension to time-varying data is challenging. We present a novel solution based on continuous scatterplot~(CSP)~\cite{Bachthaler2008CSP} and CSP operators to enable time-dependent queries on bivariate data. The CSP peel operator introduced by Sharma \etal~\cite{Sharma2023} helps analyze the bivariate field restricted to an individual spatial region, called a \emph{segment}. A direct application of the CSP peel operator to study a time-varying bivariate field requires manual inspection of CSPs of all segments at every time step, and may not be effective. We extend the CSP peel operator to provide a summary representation of the time-varying bivariate field and hence avoid the need for manual inspection of individual peeled CSPs.}

\myparagraph{\newtextF{Summary representation.}} \newtextF{\autoref{fig:teaser} illustrates the study of evolution of electronic excited states following the absorption of light in a simple molecule using our proposed visual analysis pipeline. Herein, we study the behavior of two excited states over a period of $36$~femtoseconds~(fs) by measuring a pair of scalar fields ($\phi_h$, $\phi_p$) that represent the electronic transition density. The molecular geometry varies over time and affects the two scalar fields. The CSPs correspond to the bivariate field restricted to two segments, the C3~(orange) atom in State~1 and C4~(green) atom in State~2. The CSPs evolve and align with each other at $t=30$~fs. The CSPs corresponding to the bivariate field restricted to C3 and C4 are computed, represented in 4D space using image moments~\cite{Hu1962moments}, and projected on to the plane via a PCA plot. Time tracks of the two CSPs are analyzed via visual comparison. Points $C, D',$ and $E'$ are located near each other and have similar CSPs. Also, there is a clear visual similarity between the two tracks $A$ to $E$ and $A'$ to $E'$ -- starting at $A$/$A'$, reversing direction at $B$/$B'$, turning at $C$/$C'$, primarily following PC$_3$, making another turn at $D$/$D'$, and finally ending at $E$/$E'$. It suggests that the evolution is similar in both states. Note that the annotated CSPs at time steps $A$ and $A'$ or $B$ and $B'$ do not look similar and a manual analysis of the CSPs may not help identify the similarity in the evolution unless the CSPs themselves exhibit visual similarity. The similarity is observed from a summary representation of the entire time series. The tracks of atoms in the PCA plot provide useful insights into the evolution of the molecular system, enabling chemists to identify patterns and use domain expertise to interpret the underlying physical phenomena}.

\subsection{Related  Work}
\begin{figure}[!t]
    \centering
   \includegraphics[width=0.5\textwidth]{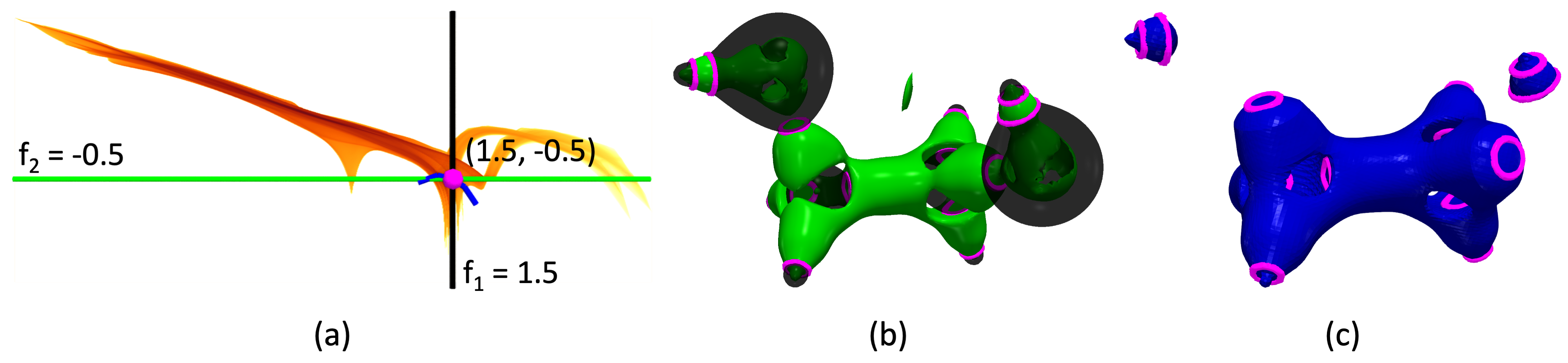}
    \caption{Fibers and fiber surface. (a)~The CSP of a bivariate field ($f_1$: electron density, $f_2$: reduced gradient) defined on an ethane-diol molecule. The black vertical line represents an isovalue for $f_1$, and the green horizontal line represents an isovalue for $f_2$. They intersect at the pink point ($1.5, -0.5$). (b)~Corresponding isosurfaces. The green isosurface encloses regions corresponding to both covalent and non-covalent bonds and some individual atoms. The two isosurfaces intersect along the pink fiber. (c)~The fiber surface (blue), corresponding to the blue control polygon selected in the CSP, highlights the covalent bonds.} 
\vspace{-0.4em}
    \label{fig:backGroundFS}
\end{figure}

\myparagraph{Bivariate field analysis.}
Continuous scatterplots~\cite{Bachthaler2008CSP} and fiber surfaces~\cite{carr2015fiber} have proven to be effective tools for analyzing bivariate fields. CSP is a generalization of the discrete scatterplot to continuous bivariate fields defined on an $n$-dimensional domain. It captures and visually presents the relationship between the two individual fields under consideration. A study of this relationship helps in the identification of interesting isovalues~\cite{nagaraj2010relation} and regions in the bivariate range space~\cite{carr2015fiber}. The scatterplot is a highly popular visual representation of bivariate data. It has been well studied with respect to the design decisions, data characteristics, and the supported analysis tasks~\cite{sarikaya2017scatterplots}. A \textit{fiber} is a generalization of isosurfaces to bivariate fields and is defined as the preimage of a value in the range space of the bivariate field. A \textit{fiber surface} is a collection of fibers, the preimage of a curve or a closed polygon in the range space. Global and local relationships between the two scalar fields may manifest as patterns or regions within the CSP.  The regions may be mapped back onto the spatial domain via fiber surfaces to study the relationship further. Carr~\etal~\cite{carr2015fiber} introduced fiber surfaces and demonstrated their effective use in the study of molecular structure, by extracting bonds and atoms in a molecule using the electron density field and other derived fields as shown in~\autoref{fig:backGroundFS}. Klacansky~\etal~\cite{klacansky2016fast} described a fast algorithm for computing fiber surfaces with guaranteed accuracy. 

\myparagraph{\newtext{Applications.}}
CSP and fiber surfaces have been shown to be useful in the study of bivariate and multivariate fields from multiple application domains~\newtext{\cite{He2019MultivariateSurvey}}.
Tierny \etal~\cite{tierny2016jacobi} used the Jacobi set to compute Jacobi fiber surfaces, which segment the domain based on the Reeb space~\cite{edelsbrunner2008reeb}. The Reeb space based segmentation of the spatial domain helps compute peeled CSP layers corresponding to individual segments in the domain. The individual layers enable the visualization of flow simulation data and the visualization of chemical interactions in a water dimer. Blecha~\etal~\cite{Blecha2019Nuclear} studied the effect of a nuclear waste repository based on data obtained from a multiphysics simulation. Their analysis was based on the fiber surfaces extracted from a subset of variables available from the simulation.  Raith~\etal~\cite{Raith2019Tensor} demonstrate the application of fiber surface to the study of stress fields that arise in component design in mechanical engineering. Zheng \etal~\cite{Zheng2019} use a dual representation that considers both spatial domain and range space of a bivariate field to identify equivalent regions, segmented by silhouette and boundary curves. The approach may help identify interesting segments and establish equivalence between segments from different time steps. \newtext{Fiber surfaces have also been studied with a focus on efficient computation using ray casting~\cite{Wu2017FiberRayCasting} and generalization to multivariate data~\cite{Blecha2020FSManyVariables}. Athawale \etal~\cite{Athawale2023Uncertainity} study fiber uncertainty using both parametric and nonparametric noise models.}  

\myparagraph{\newtext{Feature tracking.}}
Sharma~\etal~\cite{sharma2021segmentation} introduced a CSP peel operator and showed its use to study donor/acceptor behavior of subgroups in a molecule during the electronic transition. This operator may be combined with a CSP lens operator that is applied on the range space~\cite{Sharma2023}. In this paper, we present an extension of the CSP peel operator to analyze time-varying bivariate fields \newtext{and demonstrate its applicability for studying atomic behavior. Compared to static subgroup-based analysis~\cite{Sharma2023}, the number of instances increases significantly in such a scenario}.  Various structures have been developed to support feature tracking in time-varying univariate fields~\cite{bremer2009,saikia2015,saikia2017,sridh2023,yan2021}. Nested tracking graph~\cite{Lukasczyk2017NTG,Lukasczyk2019DNTG} is a summary graph representation that supports the analysis of the evolution of superlevel set components over time.  The construction of the nested tracking graphs is based on user-specified isovalues of interest, which are used to compute the superlevel sets and their components. The superlevel sets form a hierarchy within each time step among the components that are tracked. Such a notion of hierarchy is difficult to define in the case of bivariate fields. Our proposal to employ the CSP peel operator requires an input segmentation of the spatial domain of the bivariate field. The individual segments are assumed to be available over time. However, instead of computing spatial overlaps between segments in two consecutive time steps, we study the similarity between the peeled layers of the CSP that correspond to the segments. The similarity between the CSPs of the individual segments is visualized via a mapping to a 4-dimensional moment space as shown in \autoref{fig:teaser}.

\subsection{Contributions}
\newText{We present a novel approach for visual analysis of time-varying bivariate fields to study electronic structure evolution.} The approach is directed by the objectives of identifying spatial and temporal patterns and localizing spatial regions that exhibit similar behavior. The analysis brings together a diverse set of techniques resulting in a summary representation of the time-varying bivariate field that is amenable to finer-grained analysis. Specifically, the approach is based on the CSP peel operator~\cite{Sharma2023} that was developed for \newtextF{studying electronic transitions by analyzing bivariate fields} via continuous scatterplots, image moments developed as a shape descriptor in the areas of image processing and computer vision, and dimensionality reduction via principal component analysis~(PCA). The summary representation is a collection of tracks in the PCA plot. Key time steps and segments of interest in the spatial domain of the bivariate field are identified based on subsequent visual analysis of the tracks using fiber surfaces. Primary contributions of this paper include
\begin{enumerate}
    \item An image moment-based method for visualizing and exploring the similarity and dissimilarity between \newtext{atomic CSPs that characterize electronic transitions}.
    \item A new 4D representation of a time-varying bivariate field based on shape analysis of CSPs and its visual presentation via a PCA plot.
    \item A visual analysis pipeline that employs the above methods to support the exploration of time-varying bivariate fields \newtext{towards studying electronic structure evolution}.
    \item \newText{Case studies that demonstrate the applicability and effectiveness of the visual analysis pipeline to handle a large number of time instances for studying excited-state dynamics in two molecular systems to identify the active atoms, important time steps, and any structural changes in the molecule.}   
\end{enumerate}
\newText{Specifically, the case studies show how the proposed method provides a compact representation of changes in the electronic structure along nuclear trajectories that facilitate the identification of key structural modes relevant to the dynamics. The observations from the case studies have also opened up avenues for further work toward gaining insight into the study of the dynamics of the molecular system via CSPs.}

\begin{figure*}[!t]
    \centering
   \includegraphics[width=\textwidth]{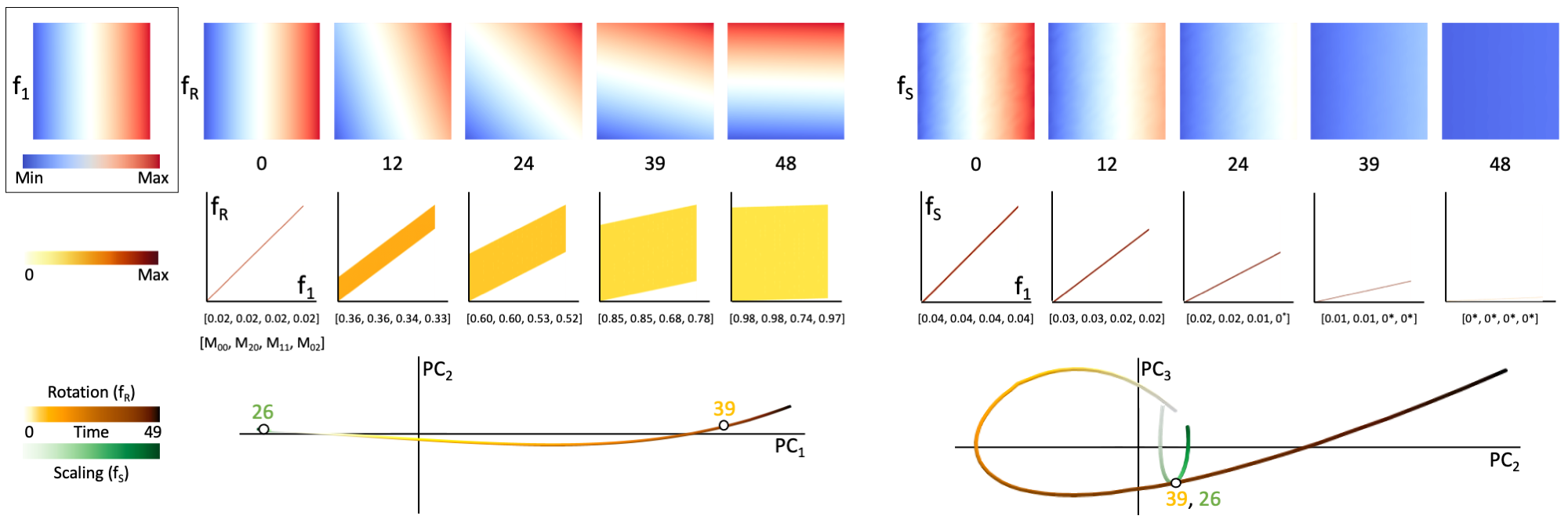}
    \caption{CSPs and tracks in the PCA plot of synthetic time-varying bivariate fields. (Top)~Two synthetic time-varying bivariate fields $(f_1,f_R)$ and $(f_1,f_S)$, where $f_1(x,y) = x$ and $f_R, f_S$ are rotated and scaled version of $f_1$. The angle of rotation and the scale factor increases with time (sampled at $0,12,24,39,48$). (Middle)~CSPs of the two bivariate fields $(f_1, f_R)$ and $(f_1, f_S)$ at each time step. An increase in rotation angle causes the CSP area to increase, and a decrease in scale factor causes the slope to decrease over time. CSPs are annotated with corresponding normalized moment vectors. Values smaller than 0.01 are marked with *. All CSPs in this paper are shown using a yellows (\includegraphics[width=0.07\textwidth]{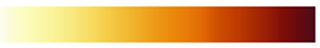}) color map in log scale. (Bottom)~Tracks of the two time-varying bivariate fields begin close to each other in both PCA plots (PC$_1$-PC$_2$ and PC$_2$-PC$_3$ space), indicating similarity between the CSPs during the initial time steps. The larger range covered by the moments for $(f_1,f_R)$ is better depicted in the PC$_1$-PC$_2$ space. The plot in the PC$_2$-PC$_3$ space provides good insight into the similar evolution pattern of the two tracks.}
\vspace{-0.5em}
    \label{fig:backGround}
\end{figure*}

\begin{figure*}[!t]
    \centering
   \includegraphics[width=\textwidth]{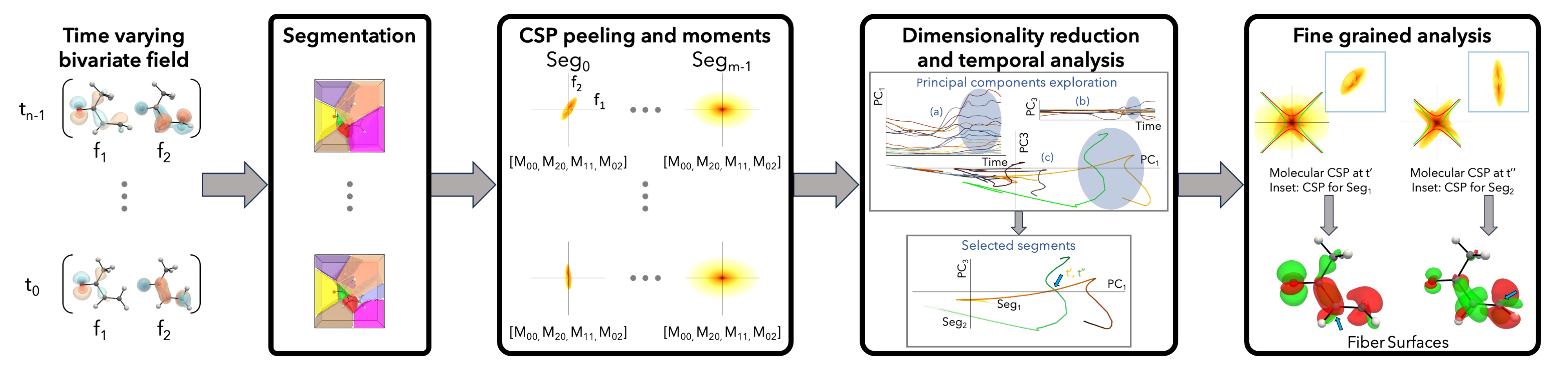}
    \caption{Visual analysis pipeline. The input consists of a time-varying bivariate field together with domain-specific information regarding potential regions of interest. The first step computes a segmentation of the domain, either using a topological or geometric approach, using the domain-specific information provided with the input. Next, the CSP and moment vectors are computed for all the segments for all time steps. In the following step, the 4D moment vectors are projected to the plane using standard dimensionality reduction techniques. Different projections and the tracks within each PCA plot are analyzed visually to identify interesting tracks. The selected tracks can be further analyzed at a finer level of granularity by studying the individual CSPs and performing comparative analysis. The blue arrow in the PCA plot highlights the intersection point of the selected tracks. The fine-grained analysis (insets) shows a similarity in the CSPs of segments at the intersection point. The fiber surfaces (red and green) in the vicinity of the corresponding atoms (indicated by the blue arrow) further confirm the observation.} 
\vspace{-0.5em}
    \label{fig:pipeline}
\end{figure*}

\section{Bivariate field analysis}\label{sec:background}
In this section, we introduce the necessary terms and mathematical structures to describe our proposed analysis methodology.

\myparagraph{Bivariate field.} A \emph{bivariate field} $f$ is a special case of a multivariate field that consists of two scalar fields defined on a spatial domain $\mathcal{D}$.
\begin{equation}
f = \{f_1,f_2\} : \mathcal{D} \rightarrow \mathbb{R}^2.
\end{equation}
Studying the two individual fields as a bivariate field allows for the identification and exploration of features and characteristics of the relationship between the two fields. If one or both of the individual fields change over time, we have a 1-parameter family of bivariate fields, called a \emph{time-varying bivariate field}. The time-varying field(s) may result in a dynamically changing relationship between the two fields. \newtext{To illustrate the concept,} \autoref{fig:backGround} shows two 2D time-varying fields.  In both bivariate fields, the first field $f_1 (x,y) = x$, is fixed over time. The second field is obtained by applying a transformation on $f_1$. The field $f_R$ is obtained by rotating about the origin by an angle that is proportional to time:
\begin{equation}
f_R (x,y;t) = a_t \cdot y + (1- a_t) \cdot f_1 (x,y).
\end{equation}
The coefficient $a_t$ initialized to the value $0.01$ at $t=0$ and incremented by $0.02$ at every integral time step. The field $f_S$ is computed by scaling $f_1$ with a scale factor that is proportion to time:
\begin{equation}
f_S (x,y;t) = a_t \cdot (f_1 (x,y) + b).
\end{equation}
Here, $b$ is a random number between $-0.5$ and $0$, $a_t$ is initialized to the value $1$ and decremented by $-0.02$ at every integral time step.
Both $f_R$ and $f_S$ are computed for $50$ time steps ($t=0$ to $t=49$). \autoref{fig:backGround} shows the CSP of the bivariate fields at a few selected time steps. \newText{Unlike the synthetic bivariate fields described in this section for introducing the terminology, real bivariate fields undergo complex changes over time and not necessarily simple scaling or rotation.}

\myparagraph{Continuous scatterplot.} A CSP~\cite{Bachthaler2008CSP} is used as a visual descriptor of the bivariate field. It is a generalization of a scatterplot to continuous functions and can be considered a 2D continuous histogram. A point in the CSP represents the density of a bivariate value, which may be visualized using a suitable color map. In this paper, CSP is used as the primary tool to describe the behavior of a bivariate field and helps visualize distinct and interesting features of the data distribution. Tracking the CSP and the features within over time helps visualize the evolution of the relationship over time. 
In \autoref{fig:backGround}, we observe that the area covered by the CSPs of $(f_1, f_R)$ increases with the angle of rotation/time. The area attains a maximum at time $t'$ when $f_R (x,y,t') = y$. The CSP aligns with the $x = y$ line at $t=0$, and the slope decreases with time and increasing scale factor.

\myparagraph{Image moments.}\label{sec:imageMomentsBg} 
\emph{Image moments}~\cite{Hu1962moments} is a shape descriptor widely used in computer vision and image processing applications. The moments capture characteristic properties of an image, such as area, centroid, and orientation. The moments are computed as a weighted average of the image pixel intensities. For a continuous 2D function $f$ and $p,q = 0,1,2,3,\ldots$, the raw moments of order-($p+q$) are defined as 
\begin{equation}
M_{pq}=\int\limits_{-\infty}^{\infty} \int\limits_{-\infty}^{\infty} x^py^qf(x,y) \,dx\, dy.
\end{equation}
The raw moments for a discrete grayscale image are defined as
\begin{equation}
M_{ij} = \sum_x \sum_y x^i y^j I(x,y),
\end{equation}
where $I(x,y)$ denotes the pixel intensities. The uniqueness theorem by Hu \etal~\cite{Hu1962moments} states that if $f(x,y)$ is piecewise continuous and nonzero only in a finite area of the $xy$ plane, then moments of all orders exist and the moment sequence ($M_{pq}$) is uniquely determined by $f(x,y)$. Hence, the moment vector can serve as a unique descriptor of $f(x,y)$ or the input image. In \autoref{fig:backGround}, the CSPs are annotated with moments vectors $[M_{00}, M_{20}, M_{11}, M_{02}]$. The moments are normalized across all time steps. Moment computation and normalization specific to CSP are discussed in \autoref{sec:momentsForCSP}. The moment $M_{00}$ captures the sum of pixel densities or area of the image. The order-2 moments capture the covariance of data and can be used to compute principal eigenvectors or the orientation. We include the order-0 and order-2 raw moments into a 4D vector to describe the CSP. In \autoref{fig:backGround}, both $(f_1, f_R)$ and $(f_1, f_S)$ have similar CSPs at $t=0$. Hence, their moment vectors are also similar. Over time, the area of CSPs increases for $f_R$ and decreases for $f_S$. The moment vectors capture these changes. For $f_R$, the individual moment values increase over time. For $f_S$, the values almost vanish during the later time steps. 

\myparagraph{PCA plot and track.} The set of all moment vectors can be visualized via projection onto the plane using a PCA plot. Each CSP corresponds to a point in the PCA plot. The CSPs change continuously over time, so the projected points form a \emph{track} on the plane. A track represents the evolution of the CSP and the corresponding bivariate field. In \autoref{fig:backGround}, the third row shows two different views of tracks corresponding to $(f_1,f_R)$ and $(f_1,f_S)$. The tracks are shown using different hues. In the PC$_1$-PC$_2$ plot, the track of $(f_1,f_S)$ (green) looks short in comparison to that of $(f_1,f_R)$ (yellow). Both begin from nearby locations, but the yellow track extends over a larger range in the projected space, as expected due to the large values of the moments. The evolution pattern in the green track is not visible in this view. The PC$_2$-PC$_3$ plot\newtext{, where PC$_i$ represents the $i^{th}$ principal component,} gives a better insight into the evolution pattern for both tracks. It appears that the tracks touch each other at $t=39$~(yellow track) and $t=26$~(green track). However, this is an artifact of the chosen PCA axes. In the PC$_1$-PC$_2$ plot, it is clearly visible that the tracks are close to each other only during the initial time steps. Since the tracks highly depend on chosen principal components, a careful exploration of the principal components is required during the analysis. The projection methodology as employed in our visual analysis pipeline is discussed in \autoref{sec:dimReduction}.
All tracks in the paper are shown using the same linear color map unless the time is one of the axes.

\myparagraph{Fiber.} A \emph{fiber} is defined as the set of points in $\mathcal{D}$ which map to a constant bivariate value $(f_1 = k_1, f_2 = k_2)$. A fiber is the bivariate analog of an isosurface. 
\autoref{fig:backGroundFS}(a) shows the CSP of the bivariate field ($f_1$: electron density, $f_2$ : reduced gradient)~\cite{Johnson2010JACS} for the ethane-diol molecule. The horizontal green line $f_2=-0.5$ corresponds to the green isosurface shown in \autoref{fig:backGroundFS}(b). The vertical black line $f_1=1.5$ maps to the black isosurface, highlighting different atoms of the molecule. The pink point ($f_1=1.5, f_2=-0.5$) where both the lines intersect in \autoref{fig:backGroundFS}(a) maps to the pink-colored fiber in the spatial domain. This fiber is the curve where the two corresponding isosurfaces intersect and may consist of multiple connected components.

\myparagraph{Fiber surface.} A fiber corresponds to a single bivariate value or a single point in the range space of the bivariate field. The preimage of a collection of points, namely a polygon in the range space, is a collection of fibers that form the \emph{fiber surface}. The polygon in the range space is called the \emph{fiber surface control polygon}. In \autoref{fig:backGroundFS}(b), the green isosurface highlights different bonds in the molecule and some of the atoms. In order to extract the covalent bonds, we select the blue-colored control polygon passing through the pink point. The fiber surface mapped to the selected control polygon is shown in \autoref{fig:backGroundFS}(c). The fiber surface passes through the fibers corresponding to the pink point as expected.

\section{Visual analysis pipeline} \label{sec:pipeline}
In this section, we present our proposed visual analysis pipeline (\autoref{fig:pipeline}) and describe individual methods that constitute the pipeline. Our goal is to study the behavior over time of segments that represent atoms in a bivariate field. \newtextF{We first describe the segmentation of the molecule under study into atoms~(\autoref{sec:segmentationPipeline}). Next, we describe the CSP peel operator and introduce a 4D representation of CSPs of segmented atoms using image moments~(\autoref{sec:momentsForCSP}). Subsequently, we introduce the dimensionality reduction technique used to project the image moment-based descriptors for CSPs~(\autoref{sec:dimReduction}), enabling a summary representation that visualizes the tracks corresponding to each atom~(\autoref{fig:teaser}). Finally, we discuss fine-grained analysis following a study of coarse-grained behavior using the summary representation~(\autoref{sec:fineAnalysis}).}

The illustrations in \autoref{fig:pipeline} explain the application of the pipeline on a molecular system. The time-varying bivariate field, available at $n$ time steps, and any domain-specific information is provided as input. In \autoref{fig:pipeline}, the input consists of two different excitation states of a molecule. Details of the state are case study specific, and do not influence the method description. 
The visual analysis pipeline includes \textit{browsing} and \textit{aggregate-level tasks}, following the categorization of analysis tasks applicable for scatterplots introduced by Sarikaya and Gleicher~\cite{sarikaya2017scatterplots}. The peel operator allows for exploring (browsing) specific segments or neighborhoods. Studying specific peeled CSPs and their evolution, identifying outliers, and identifying similarity between peeled CSPs are aggregate-level tasks. 

\subsection{Segmentation}\label{sec:segmentationPipeline}
\newtextF{We follow previous approaches~\cite{masood2021visual, Sharma2023} and compute a weighted Voronoi tessellation~\cite{Aurenhammer1987powerdiag} to partition the molecule into atomic regions.} The segmentation is application-dependent and depends on the features that we wish to analyze over time. \newtext{The goal in this paper is the study of atomic behavior and hence the choice of segmentation method. A segmentation is not required when studying the complete bivariate field.} The segments can be computed using a geometric or topological approach at every time step. The segments may be fixed or vary over time. For example, if the segments are determined by the vicinity to an atom in the molecule and if atom locations do not change over time, then we have a consistent geometric segment across all time steps for every atom. But, if the molecule geometry changes over time, then the segments will change as well. In any case, the bivariate field is restricted to the segment with the aim of capturing the properties of the corresponding atom. In \autoref{fig:pipeline}, the segmentation \newtextF{(weighted Voronoi tessellation~\cite{Aurenhammer1987powerdiag})} is based on the geometric location of atoms to support the analysis of atom behavior. The molecule geometry changes over time, resulting in time-varying segments. Segmentation is the first and critical step in the pipeline. \newText{The summary representation computed in subsequent steps may have unpredictable discontinuities if the segment does not represent well-defined features, for example if the bivariate field within the segment is not continuous over time}.

\subsection{CSP peeling and image moments }\label{sec:momentsForCSP}
\newtextF{The CSP peel operator is used to compute the CSPs corresponding to individual segments and image moments are used to represent each CSP as a feature vector.}
The CSP peel operator~\cite{Sharma2023} generates the CSP of the bivariate field restricted to individual segments. Since the bivariate field varies over time, the peeled CSPs capture the evolving structure of the bivariate field restricted to each segment. Visual inspection of the peeled CSPs aids in understanding the behavior of a particular segment over time and in studying similarities between different segments. Outliers may also be identified visually. However, as the number of time steps and segments increase, visual inspection is not an effective strategy. It is not feasible to perform a detailed analysis of all peeled CSPs. A CSP may be represented by a feature vector that captures its shape and density distribution, thereby mapping it to a point in a high-dimensional space, where the dimension depends on the length of the feature vector. The feature vector should serve as a good descriptor. Essentially, we desire that the feature vectors corresponding to two similar CSPs are close to each other and they are further apart if the CSPs are dissimilar. Image moments~\cite{Hu1962moments} have proven to be very useful in describing shapes and satisfy our requirements. \newtextF{The image moments are computed on the $\log$ of the density field to be consistent with the log-scale used to render the CSPs. If $d(x,y)$ is the density at a point $(x,y)$ in the CSP, then the moments are computed as
\begin{equation}
M_{ij} = \sum_x \sum_y x^i y^j \log (1+d(x,y)).
\label{eqn:CSPMomentsCompute}
\end{equation}
The shift by 1 before taking the $\log$ effectively handles points with zero density. The logarithmic scaling is not required in the moment computation if the CSPs exhibit distinct patterns without a log-scale mapping.} 

We use a finite number of image moments, up to an appropriate order, depending on the shape characteristics that we desire to be captured. The order-0 moment captures the area of the image, as described in \autoref{sec:background}. For CSPs, it captures the sum of densities, which can be considered a rough measure of the size of the CSP. The order-2 moments capture the covariance and orientation of the CSP. As a minimal requirement, we would like the CSPs to be characterized by the area, orientation, and distribution of density within. These properties are captured by the order-0 and order-2 moments. Specifically, the CSPs of the bivariate fields that we study in the case studies are prominently oval or spherical in shape and hence benefit from the above choice of moments. In this paper, the density at each point of the CSP is directly used to compute the moments \newtextF{(\autoref{eqn:CSPMomentsCompute})} rather than rendering CSPs as images and using the image pixel densities. This approach reduces the loss of information due to rendering the CSPs as images. The range of values of the moments of order-0 and order-2 may vary significantly.  $M_{00}$, which is the measure of the size of CSP, typically has much higher values than the other moments. So, the distance between two feature vectors is often dominated by $M_{00}$ unless their values are normalized. So, we normalize the moments so that they assume values between 0 and 1. The normalized value $M_{ij}^{N}$ of an order-$p$ moment ($p=i+j$) is
 \begin{equation}
M_{ij}^{N} = \frac{M_{ij} - \min(M_{p})}{\max(M_{p}) - \min(M_{p})}
\end{equation}
Here, $\min(M_{p})$ and $\max(M_{p})$ are the minimum and maximum values across all the CSPs. The normalization is global, it is computed for all the segments across all time steps that are considered in the analysis. \newtext{The normalization process is sensitive to outliers. However, in the intended application, this sensitivity is desirable for identifying the tracks of outlier atoms, as illustrated in \autoref{fig:CS1ImportantAtoms} (track of S1\_O). The data is generated using simulations as described in \autoref{sec:ESDT}, and is unlikely to contain outliers.
} To simplify the notation, we use $M_{ij}$ to denote the normalized moment value $M_{ij}^{N}$ in the rest of the paper. We assign equal weights for all the moments. 

Application-specific \newtext{feature descriptors} may also be incorporated in conjunction with the standard moments. For example, during an electronic transition, charge lost, and charge gained are the scalar fields that decide whether a particular \newtextF{atom} behaves as a donor or acceptor. The total amount of charge lost by a particular \newtextF{atom} is called its donor strength. Two different atoms with similar donor strength may be considered as behaving similarly. So, the donor strength can be used as an application-specific \newtext{feature descriptor}. In this paper, we consider only the standard moments to produce a feature vector of length 4, $[M_{00}, M_{20}, M_{11}, M_{02}]$, refer \autoref{sec:ESDT}.

\subsection{Dimensionality reduction and temporal analysis}\label{sec:dimReduction}
At least four moments are required to describe the size and orientation of the CSP. We use standard dimensionality reduction methods to project the moment vectors onto the plane for visualization. We experimented with different projection techniques, such as multidimensional scaling~(MDS), principal component analysis (PCA), and t-distributed stochastic neighbor embedding (t-SNE). \newText{MDS results in abrupt jumps in the track as shown in the supplementary material because it may project the same point from two adjacent time steps to distant lower dimensional points. t-SNE poses the challenge of specifying hyperparameters. Incorrect tracks generated by such a method can make it challenging to determine whether the chosen dimensionality reduction method is unsuitable or if further hyperparameter tuning is necessary.} PCA was most amenable to the constraint of maintaining continuity across adjacent time steps \newtextF{and} gives results aligning with the expected behavior of segments.

The choice of dimensionality reduction method depends on the properties to be preserved during projection. The analysis works with any technique if it yields continuous projections over time. PCA aims to preserve the data variance  \newtextF{and applying it to individual time steps preserves the variance restricted to that time step. When applied to each time step independently, the set of projected points may not be consistent and not amenable to visual analysis. So, we apply the dimensionality reduction step on all CSPs collected across all time steps. The projection may be sliced into projections corresponding to individual time steps or segments and visualized separately. When sliced into PCA of individual segments, the tracks corresponding to a particular segment are displayed as piecewise linear curves.}

Choosing the top two principal components results in a plot that explains the large variance, but application-specific features and patterns may be lost. \newtext{The individual principal components assign lower weight to one or more of the four image moments.} If the desired properties span very few segments in the domain or a small number of time steps, then PCA may ignore them, and the patterns may not be visible due to visual clutter. \newtext{Therefore, it is essential to have a combination that captures all four moments to some extent in order to accurately represent the size and orientation of CSPs and to summarize the evolution of individual atoms in a lower-dimensional space.} To address this issue, we explore the principal components individually and pairwise with the aim of locating interesting patterns. 

In \autoref{fig:pipeline}, under principal components exploration, the 1D tracks of individual segments for PC$_1$ and PC$_3$ exhibit some distinct patterns in the later time steps. Further, projecting the segments in the PC$_1$-PC$_3$ space, we observe a similar pattern of evolution between the green and yellow tracks. Translucent ovals highlight the patterns. The tracks can be visually analyzed to identify outliers, any consistent patterns across all tracks, and to identify segments that are not very active over time. Tracks, that cover a small area, correspond to the segments whose peeled CSPs do not change significantly over time in terms of area and orientation. \newText{The turns in the tracks may correspond to the time steps where the CSP evolution pattern changes significantly. Such time steps are likely to denote the time steps of interest.}  The visual analysis helps to identify or select segments to be analyzed in detail. In \autoref{fig:pipeline}, the tracks for two segments are selected. Both intersect at a point highlighted by the blue arrow at time step $t^\prime$ for segment $Seg_1$ and $t^{\prime\prime}$ for $Seg_2$. \newText{The intersection suggests a likelihood of similarity between the two segments.} Each step of the analysis pipeline applies a transformation or change of representation that may introduce a loss of information. This implies that an intersection between two tracks or the appearance of two tracks close to each other is an indication of similarity but does not guarantee the similarity between the CSPs. Such occurrences indicate that the relevant time steps and CSPs need to be analyzed further.

\subsection{Fine-grained analysis}\label{sec:fineAnalysis}
\newtextF{In the final step of the pipeline, we study CSPs of segments and time steps of interest as a part of a fine-grained analysis. Plotting appropriate fiber surfaces also helps in the study.} The tracks in the summary representation allow for coarse-grained analysis and observations regarding the segments and their evolution. Observing their track length enables us to identify segments that do not change much over time. The points of intersection between tracks help identify time steps when two or more segments may exhibit similar behavior. 

In \autoref{fig:pipeline}, the fine-grained analysis stage of the pipeline shows the CSPs of the molecule at the specific time steps where the two tracks intersect. The inset shows the peeled CSPs restricted to the corresponding segments. The peeled CSPs do not appear to be equal but display some similarity. The CSP for $Seg_2$ is vertically aligned, whereas the CSP of $Seg_1$ exhibits some variation along the vertical axis. To analyze the CSPs further, we trace consistent red and green control polygons in both CSPs. The corresponding fiber surfaces indicate similar red and green fiber surfaces encapsulating the atoms (highlighted by blue arrows). The similarity between the fiber surfaces indicates that both atoms behave similarly at the time steps where the two tracks intersect. 
\begin{figure}[!t]
    \centering
   \includegraphics[width=0.35\textwidth]{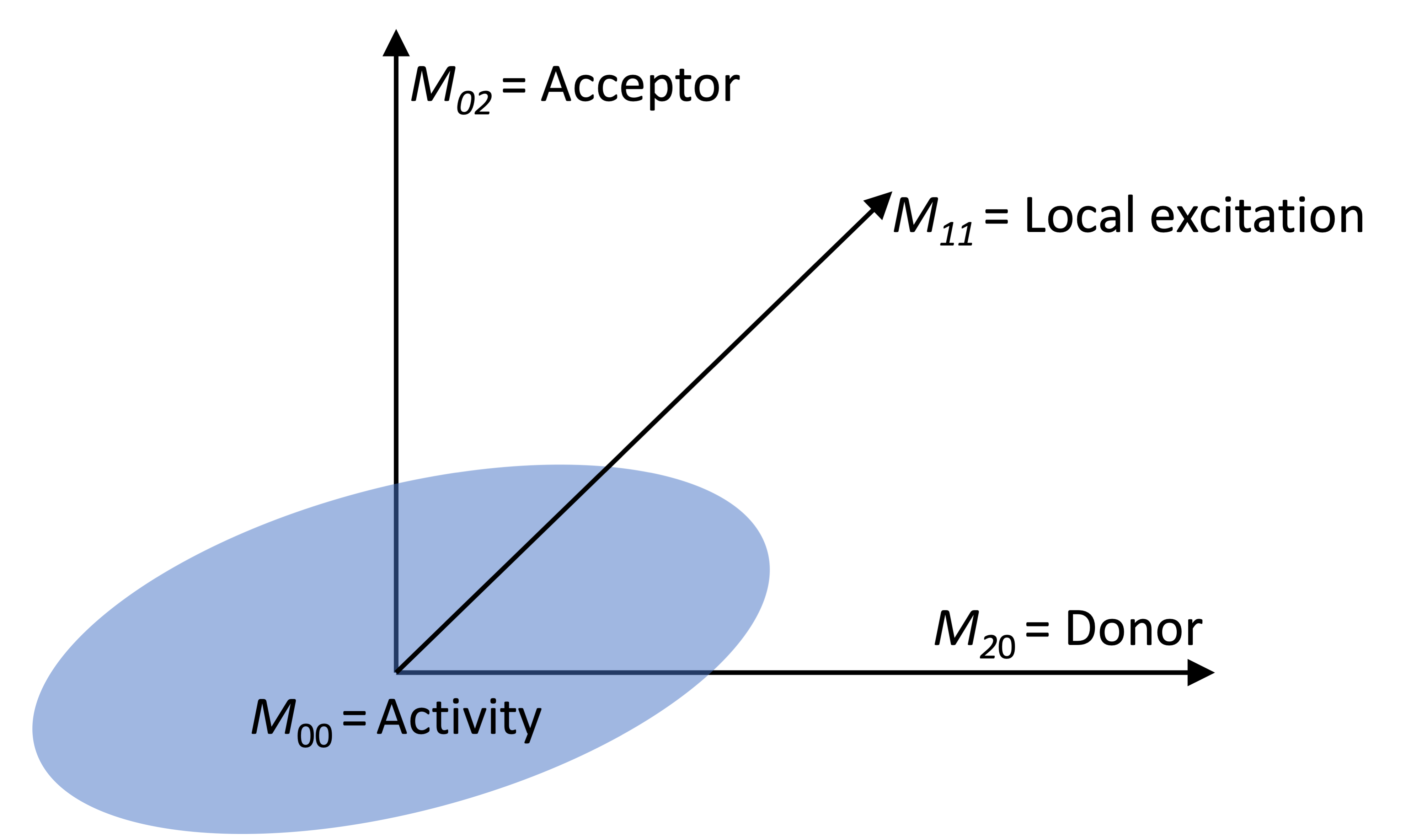}
    \caption{\newText{Image moments to describe atom's CSP (blue oval). $M_{00}$: captures the area, larger area implies larger activity. $M_{20}$ captures variation along horizontal hole\_NTO axis or donor character, $M_{02}$ captures acceptor behavior, and a diagonal CSP signifies local movement of charge.}}
\vspace{-0.4em}
    \label{fig:momentChoice}
\end{figure}

\section{\newText{Excited-state dynamics trajectories}}\label{sec:ESDT}
\newtextF{This section describes the objectives in the study of excited-state dynamics, introduces the molecules that are analyzed, and presents a breakdown of the study into visual analysis tasks.}
Light-driven dynamics is of central importance in both natural and engineered systems where the energy from light is used to trigger and drive a multitude of processes. The photon energy promotes the molecule into an electronic excited state, whose potential energy surface and network of couplings with other states will determine the fate of the nuclear dynamics and, ultimately, how the photon energy will be transformed. To decipher the inner workings of these processes, it is therefore critical to be able to characterize and understand the evolving electronic and nuclear structure. Here, we analyze the electronic changes along single representative trajectories (that is, evolving nuclear geometries) obtained from excited-state dynamics simulations of gas-phase methylvinylketone (MVK) and \textit{cis}-stilbene. The former is a major atmospheric oxidation product of biogenic isoprene, whereas the latter is a prototype photoisomerization system. The data was generated using \textit{ab initio} multiple spawning (AIMS\cite{AIMS_bennun}) simulations with the hh-TDA\cite{bannwarth_hhtda_2022} method (MVK: $\omega$PBEh/6-31G**\cite{chakraborty2023} and \textit{cis}-stilbene: BH\&HLYP/def2-TZVP).

To characterize the changes in electronic structure along the evolving nuclear geometries, we use a natural transition orbital (NTO) decomposition\cite{martin_nto_2023}. This provides a compact representation of the one-electron transition density in terms of hole and particle orbital pairs. The hole and particle NTOs  (labeled $\phi_h$ and $\phi_p$, respectively) represent the electronic change (from where and to where the electron is moved in the transition, respectively) for a given electronic excited state relative to the ground state. Specifically, we consider the temporal evolution of the dominant NTO pair for each electronic excited state as a time-varying bivariate field. The associated CSP provides a visual representation of how the spatial distribution of the bivariate field changes over time. The horizontal and vertical axes of CSPs represent hole and particle NTOs, respectively. Hence, a horizontally aligned CSP implies that the atom is a strong acceptor, a vertically aligned CSP implies a strong donor and a diagonal CSP implies that the atom exhibits both donor and acceptor character \ie local excitation~\cite{sharma2021segmentation}. As shown in \autoref{fig:momentChoice}, the three properties, donor, acceptor, and local excitation, are captured by the moments $M_{20}, M_{02}$ and $M_{11}$ respectively. The size of CSP captured by $M_{00}$ indicates how active the atom is. Hence, a vector consisting of these four moments is used to describe the CSPs generated using ($\phi_h$, $\phi_p$) bivariate field. Considering other moments may capture fine details of the CSP but impacts these prominent features of interest while projecting to lower dimensions. \newtext{Refer to Section 2 in the supplementary material for the variation of moments over time.} The study of the excited-state dynamics requires the following tasks: 
\begin{enumerate}
    \item[T1.] Identify outliers and active/inactive atoms.
    \item[T2.] Identify the time steps that correspond to a distinct atom behavior.
    \item[T3.] Identify and analyze similar or opposite patterns of evolution for active atoms.
    \item[T4.] Identify structural changes in the bonds caused by changes in atom locations.   
\end{enumerate}

\begin{figure}[!t]
    \centering
   \includegraphics[width=\columnwidth]{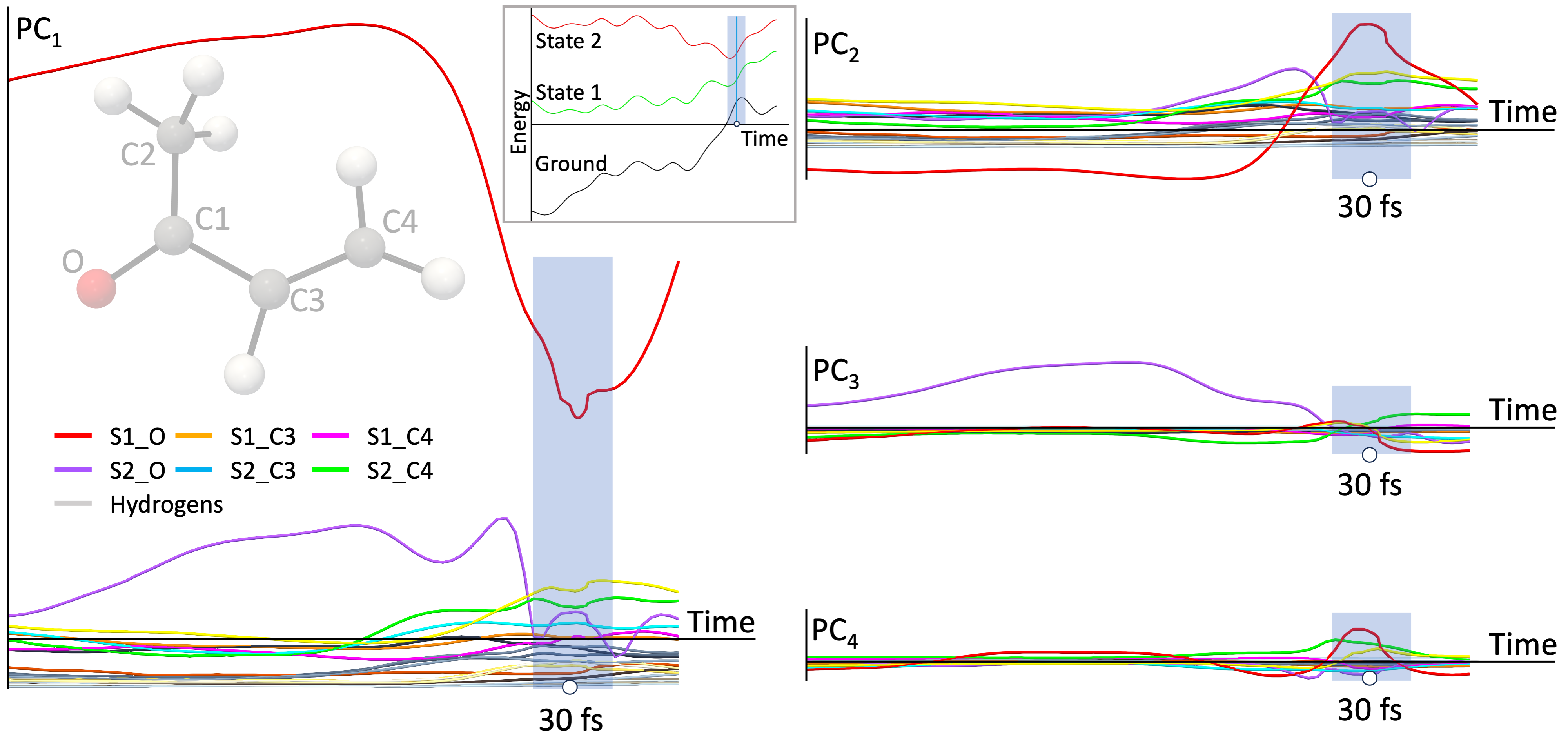}
    \caption{Analyzing individual principal components for the molecule MVK. Individual carbons in the molecule are indexed for ease of reference, hydrogen atoms are not labeled. Each track corresponds to an individual atom. Atoms that are not active during the dynamics are shown in gray color. We observe high activity within the PC$_1$ plot. The oxygen atom exhibits a distinct behavior compared to other atoms. S1\_O appears to be an outlier. The blue rectangle highlights a time interval containing interesting patterns in all the principal components.} 
\vspace{-0.5em}
    \label{fig:CS1PCAnalysis}
\end{figure}

\newText{At nuclear geometries where the energy of two electronic states crosses (so-called \emph{conical intersection seam}), we expect a strong coupling between the electronic states and hence mixing and interchange of their characters. Such regions are approached at ${\sim}30$~femtoseconds (fs) as highlighted by the blue translucent rectangle in the energy diagram, \autoref{fig:CS1PCAnalysis}. AIMS uses adaptive time stepping \ie the time step is reduced in the vicinity of the intersection seam.}

\section{Results}\label{sec:results}
We present two case studies on data from photoinduced molecular dynamics simulations to demonstrate our bivariate analysis method. In this section, we introduce the application, mention the objectives of the simulation study, describe the data, and present two case studies. \newtext{The method development, case study identification, and interpretations were conducted in close collaboration with a theoretical chemist and a coauthor of this paper.} We use Paraview~\cite{Ayachit2015paraview} together with the Topology Toolkit (TTK)~\cite{Tierny2018ttk} to compute the CSPs and generate the visualizations. The principal components and associated computations are based on the PCA module from scikit-learn~\cite{scikit-learn}. Further details regarding the implementation and runtime statistics are available in the supplementary material.

\subsection{Case Study 1: Methylvinylketone} \label{sec:MVK}
In the first case study, we analyze the earliest dynamical behavior (initial $36$~fs) of methylvinylketone (molecular structure in \autoref{fig:CS1PCAnalysis}) upon excitation to S$_2(\pi\pi^*)$ (State 2). \newtext{The dynamics represent C=C double-bond photoisomerization, a prototypical photoreaction in polyenes.} Specifically, the change in electron distribution upon excitation causes rotation around the C3-C4 dihedral angle that brings the system near a conical intersection seam with the S$_1$~state (State 1)\cite{chakraborty2023,hohenstein_2021}, highlighted by the translucent blue rectangle around $30$~fs in \autoref{fig:CS1PCAnalysis}. The single trajectory considered here represents the initial adiabatic dynamics on S$_2(\pi\pi^*)$ approaching the intersection seam. 


 The molecule contains $11$ atoms, and we consider $83$ time steps (actual elapsed time is indicated in femtoseconds, fs) for the two electronic states, yielding a total of $2 \times 11 \times 82$ bivariate instances. Individual bivariate instance is segmented based on atom locations, the moments are computed and normalized as discussed in \autoref{sec:momentsForCSP}. The normalized moments are projected using PCA to compute the tracks to show the evolution of every atom for both states over all the time steps.  
\begin{figure*}[!t]
    \centering
   \includegraphics[width=\textwidth]{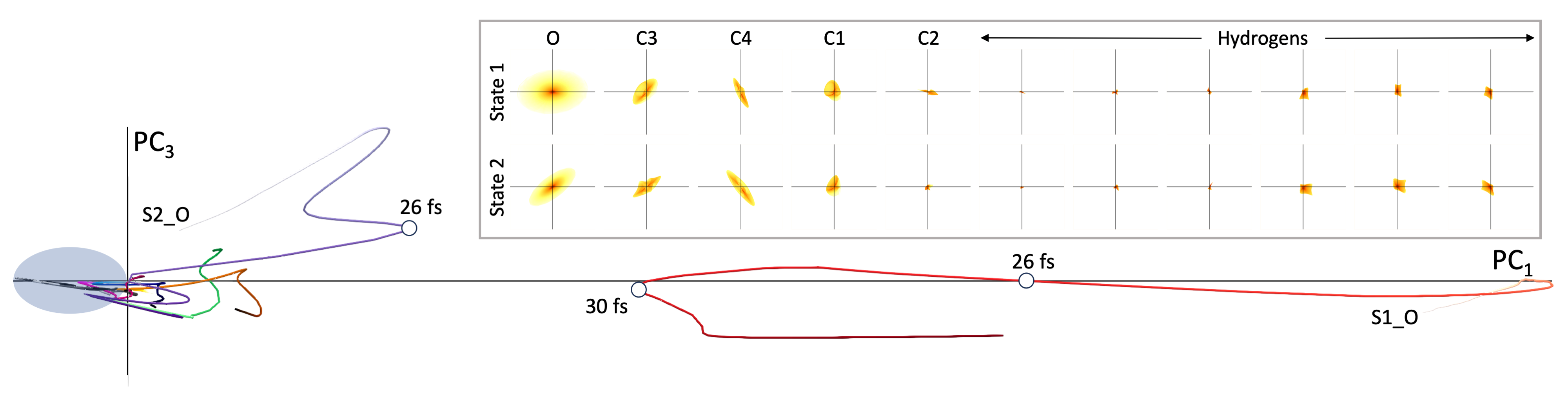}
    \caption{Analyzing track of all atoms from both states in the PC$_1$-PC$_3$ plot. Oxygen in both states (S1\_O and S2\_O) shows a distinct pattern. The track of S2\_O merges with other atoms after $26$~fs, and S1\_O remains an outlier. The inset shows that the CSP of S1\_O has a comparatively larger CSP, which explains the behavior of the track. All atoms exhibit a bulge around $30$~fs, which requires further analysis. The blue oval highlights a cluster of small tracks that seem to indicate low activity. As expected, the hydrogens and the methyl carbon C2, whose tracks appear within the oval, do not participate significantly.} 
\vspace{-0.5em}
    \label{fig:CS1ImportantAtoms}
\end{figure*}
Individual principal components are analyzed using the 1D plots shown in ~\autoref{fig:CS1PCAnalysis}. Each track corresponds to an atom.  We use the notation SX\_Y to indicate atom Y when considering electronic state X. 
To avoid clutter, only atoms (the carbonyl oxygen, O, and the vinyl carbons, C3 and C4) that undergo significant changes are colored, and the others are shown in gray. All the tracks show a distinct pattern as the system reaches a conical intersection seam around $30$~fs. A 2D plot \ie tracks computed by projecting the moments to a pair of principal components may reveal more insights. A preliminary analysis of the PCA plots, described in the supplementary material, indicates that the PC$_1$-PC$_3$ plot provides relevant application-specific insights, so we restrict the following analysis to this plot.

\newtext{In the following, we summarize the key findings from the visual analysis. The results are organized based on the tasks identified in \autoref{sec:ESDT}, namely the identification of important atoms (T1), timesteps (T2), and a study of the evolution patterns (T3) of key atoms.}

\myparagraph{Important atoms (T1).}
\autoref{fig:CS1ImportantAtoms} shows tracks in the PC$_1$-PC$_3$ space for all atoms. S1\_O is an outlier due to its large PC$_1$ component in all time steps. This outlier behavior is due to oxygen's lone-pair dominated hole NTO. S2\_O also spans a large region, but it merges with other tracks after $26$~fs, indicating that, in later time steps, its behavior is similar to other atoms. The blue oval highlights a region where all the hydrogens, C1, and C2, from both states, are clustered together. It indicates that the corresponding CSPs have a negligible area or there is no variation in the CSPs. Hence, the tracks are located within a small region, suggesting that the atoms are comparatively less active than others. 
However, the corresponding atoms may have a span along other principal components. Hence, further analysis is required. To confirm the observations with respect to atoms, CSPs corresponding to one time step ($26$~fs) are shown in the inset. The CSP for S1\_O has a larger area as compared to other atoms. Hydrogens and the two carbons (C1, C2) have CSPs with smaller areas with no strong orientation. The same behavior is observed in other time steps. We focus our attention on C3, C4, and S2\_O in the following discussion.

\begin{figure*}[!t]
    \centering
   \includegraphics[width=\textwidth]{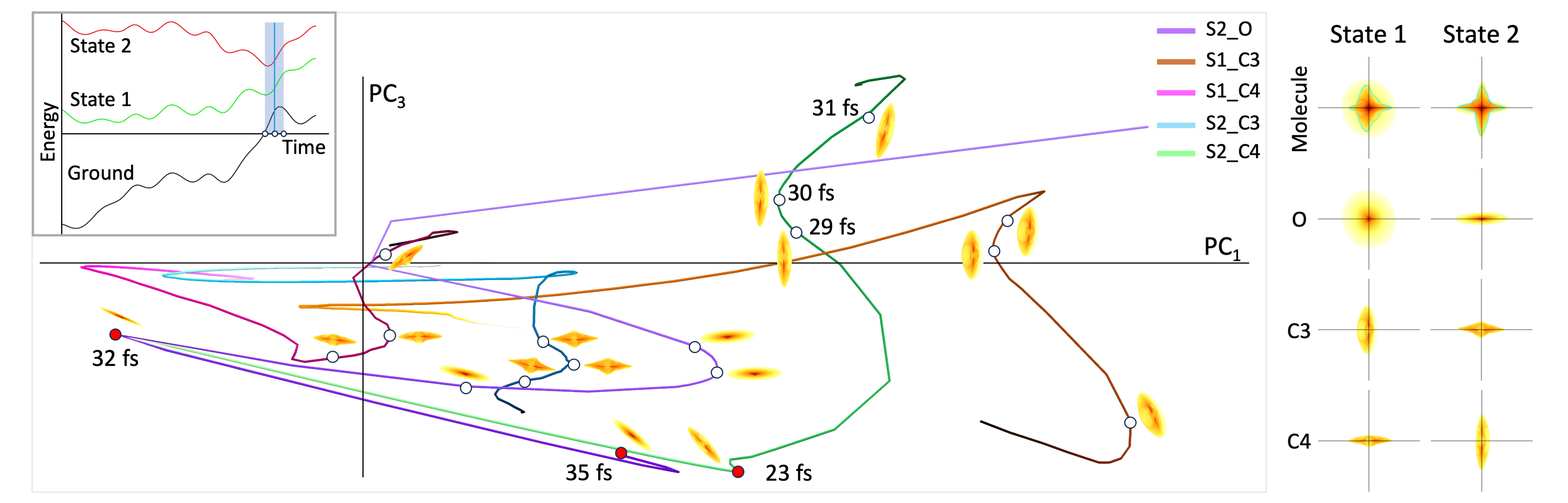}
    \caption{Analyzing the tracks of S2\_O and the two vinyl carbons C3 and C4 in both states. All tracks exhibit a uniform bulge around $30$~fs. This time window corresponds to nuclear geometries where the S$_2$ and S$_1$ states, as well as the ground state, come closer in energy (see inset, blue rectangle). All CSPs either align horizontally or vertically at $30$~fs (right), after which their orientation changes (see CSP annotations to the track). The tracks of S2\_O and S2\_C4 run parallel to each other for a short time interval (highlighted by red points). This indicates that the behavior of S2\_O towards the end of the simulation is similar to that of S2\_C4 at the start of the simulation.} 
\vspace{-0.5em}
    \label{fig:CS1ImportantTimes}
\end{figure*}

\myparagraph{Important time steps (T2).}
In \autoref{fig:CS1ImportantAtoms}, the track for S1\_O has a bulge around $30$~fs and comes closest to other tracks at this time. \autoref{fig:CS1ImportantTimes} shows that the tracks of a few selected atoms exhibit a similar bulge. While this behavior is visible in the 1D plots in \autoref{fig:CS1PCAnalysis}, the patterns are not as apparent as in the 2D plot. 
Further analysis of the CSPs near $30$~fs (see CSP annotations of tracks) reveals that all atoms begin exhibiting significant changes at $29$~fs, their CSPs align horizontally/vertically at $30$~fs, and undergo comparatively smaller changes after $31$~fs. At $30$~fs, S2\_C4 and S1\_C3 have vertically aligned CSPs (strong acceptor), whereas the other three have horizontally aligned CSPs (strong donor). Further, the tracks are well segregated during this time interval. Inspecting the CSPs, we note that the donors (S1\_C4 and S2\_C3) lie prominently towards the left. The molecular CSP for both states is aligned with the axes (observe the aqua contour). These rapid changes in the bivariate field indicate that geometries at these particular time steps may have caused some unique behavior in these atoms. Indeed, we expect that as the states get closer in terms of energy (\autoref{fig:CS1ImportantTimes}, inset), they mix and interchange characters. The three white points on the $x$-axis within the blue rectangle in the inset highlight these times. This behavior occurs near $90^{\circ}$ twisted geometries and leads to a decoupling of the $\pi$-conjugated system, in turn, producing the observed donor/acceptor behavior of the vinyl carbons C3 and C4. 
Another interesting observation from this summary representation is that the behavior of S2\_O in later time steps ($32-35$~fs) is similar to that of S2\_C4 in the early time steps ($0-23$~fs) of the simulation. The parallel sections of the two tracks are shown using red points.

\begin{figure*}[!t]
    \centering
   \includegraphics[width=\textwidth]{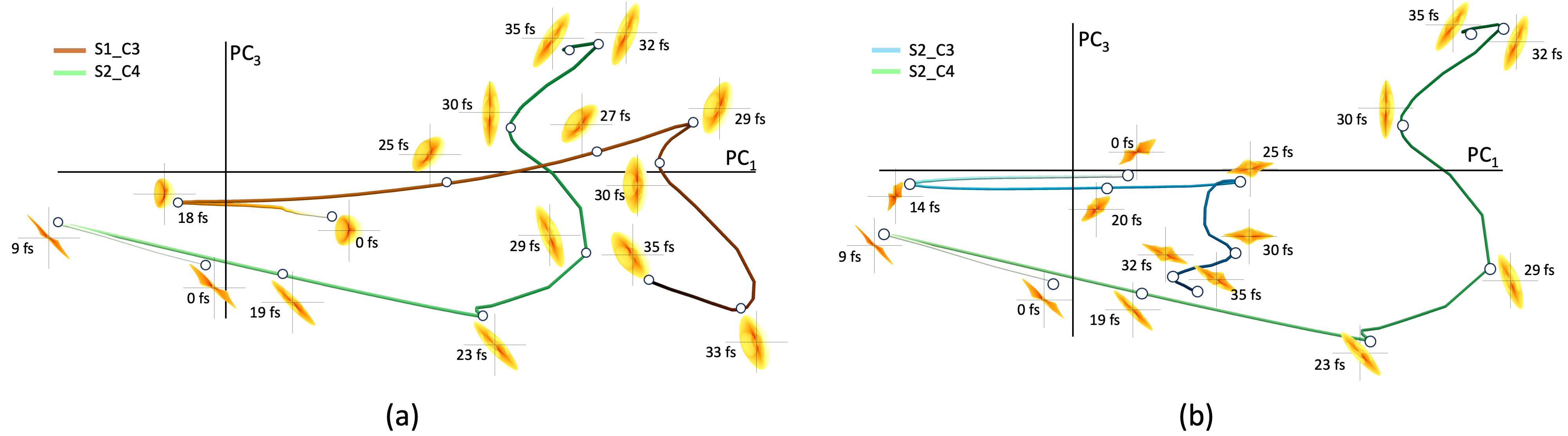}
    \caption{Analyzing the pattern of evolution between different atoms.  (a)~Across states. C3 from State S$_1$ goes through similar changes as C4 from State S$_2$ but with opposing orientations. The CSP of S1\_C3 begins as a negative diagonal at $23$~fs, rotates clockwise, and ends as a positive diagonal orientation. The CSP for S2\_C4 begins with a positive diagonal orientation at $29$~fs, rotates counterclockwise, and ends with a negative diagonal orientation. At $30$~fs, both behave as acceptors. The diagonal orientation implies charge is transferred locally, so both atoms have similar behavior overall during the time period. (b)~Within a state. Both S2\_C3 and S2\_C4 rotate clockwise. At $30$~fs, C4 behaves as a strong donor, and C3 behaves as a strong acceptor. The geometry at $30$~fs forces this change in behavior of both atoms.} 
\vspace{-0.5em}
    \label{fig:CS1carbonsEvolution}
\end{figure*}

\myparagraph{Atoms evolving similarly or in opposing directions (T2, T3).}
We observe that the tracks of S1\_C3, S1\_C4, S2\_C3, and S2\_C4 are similar in \autoref{fig:CS1ImportantTimes}. We study two scenarios in detail. \autoref{fig:CS1carbonsEvolution}(a) shows the evolution of S1\_C3 and S2\_C4. Until $23$~fs, the CSP of S2\_C4 has a negative diagonal orientation with a small change in shape. S1\_C3 does not exhibit a prominent orientation at $0$~fs. Till $29$~fs, the shape of its CSP changes gradually and acquires a positive diagonal orientation. Between $23$~fs to $35$~fs, the CSP of S2\_C4 rotates clockwise and ends in a positive diagonal orientation. In contrast, S1\_C3 rotates in the opposite direction and ends up in a negative diagonal orientation. Both become acceptors at $30$~fs. In terms of chemistry, the overall behavior of the atoms remains the same during the time interval, since both diagonal orientations denote a local movement of charge. Next, we study C3 and C4 in state S$_2$. \autoref{fig:CS1carbonsEvolution}(b) shows that their tracks are similar and the CSPs of both atoms rotate clockwise after $23$~fs. However, at $30$~fs, one becomes a strong donor while the other is a strong acceptor. This is a consequence of the decoupling of $\pi$-conjugated system occurring at $90^\circ$-twisted vinyl configurations.
\newtext{It is worth noting that the evolution pattern is not as visually prominent in \autoref{fig:CS1PCAnalysis} as it is in \autoref{fig:CS1ImportantTimes} and \autoref{fig:CS1carbonsEvolution}. This highlights the usefulness of 2D PCA plots compared to 1D PCA plots.} Additional observations regarding the evolution of S2\_O are described in the supplementary material.

\begin{figure*}[!t]
    \centering
   \includegraphics[width=\textwidth]{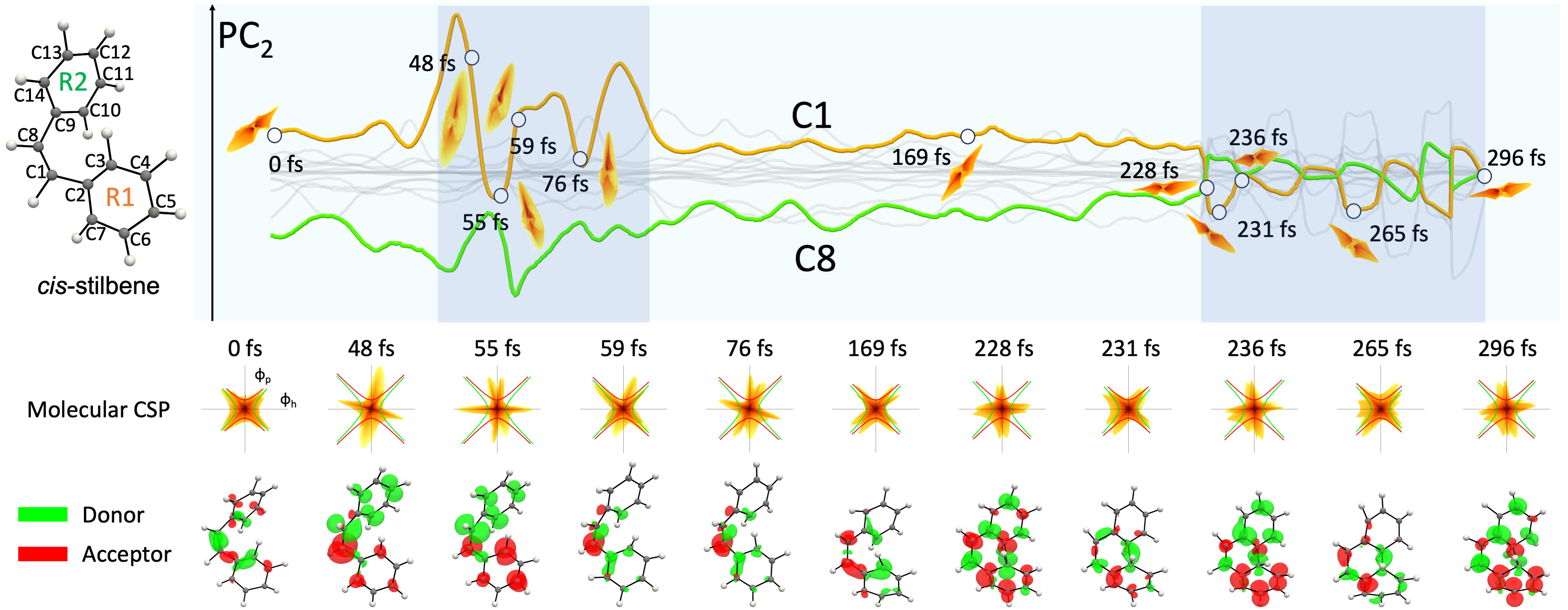}
    \caption{Evolution of the bivariate field of \textit{cis}-stilbene. (Top)~Tracks for C1 and C8 in the PC$_2$ plot. The track for C1 is annotated at a few notable time steps. The peeled CSPs for C1 typically exhibit a diagonal orientation except for time intervals where the tracks exhibit a variation in PC$_2$ (near $48$~fs and after $228$~fs). (Middle)~The molecular CSPs show a similar pattern. They are typically aligned along the diagonal. The alignment changes to the axes during the above-mentioned time. (Bottom)~Donor and acceptor fiber surfaces indicate that all atoms become strong donor/acceptor during the above-mentioned time steps. A third ring forms after $228$~fs. The alignment with the axes implies one complete ring behaves as an acceptor and the other as a donor. The new ring behaves as both, with alternating carbons showing donor/acceptor behavior.} 
\vspace{-0.5em}
    \label{fig:CS2C1Evolution}
\end{figure*}

\subsection{Case Study 2: \textit{cis}-stilbene}
We next consider the behavior of \textit{cis}-stilbene upon photoexcitation to the S$_1(\pi\pi^*)$ state, considering the first ${\sim} 300$~fs. As shown in \autoref{fig:CS2C1Evolution}, it is comprised of a central ethylenic C1=C8 bridge that connects two phenyl rings (R1: C2-C7 and R2: C9-C14). The electronic transition promotes an electron from the $\pi$-orbital (dominant hole NTO) with bonding character across the ethylenic C1-C8 bridge to the $\pi^*$-orbital (dominant particle NTO) with antibonding character. \newtext{Like in case study 1, this change can promote a C=C double bond photoisomerization reaction. However, \textit{cis}-stilbene can also undergo ring-closing as characterized by C3-C10 bond formation. We will track this bond-formation in a single ring-closing trajectory on S$_1$. The associated hole and particle NTOs represent a bivariate field.} The key nuclear coordinate changing along this trajectory is the distance between the R1 and R2 rings that brings the system toward a conical intersection seam with the ground state. This leads to a strong coupling between the S$_1$ state and the ground state in the time window 230-296~fs (blue). A smaller coupling is also observed between 48-85~fs. Our goal is to investigate the impact of the geometric changes across the two rings and the bridge atoms. 

We ignore contributions from hydrogen (unlabeled, white atoms) because they do not significantly participate in the electronic changes. 
The dataset contains $704$ time steps with sampling rates of $0.48$~fs (or $0.12$~fs when the two electronic states are close energetically).
However, the analysis is not impacted by the different sampling rates. It may play an important role in some scenarios, as discussed in the supplementary material. We segment the bivariate field ($\phi_h, \phi_p$) based on atom locations at every time step. The total number of instances to analyze after segmentation is $704 \times 14$ because only the carbon atoms are considered in the computation. The moments are computed for the CSP corresponding to each instance and fed to PCA. We select PC$_2$ and PC$_3$ as the projection space. The PCA exploration details are specified in the supplementary material. For the following discussion, we restrict our attention to the 1D tracks based on PC$_2$ to understand the behavior of bridge atoms and to analyze the formation of the new bond.

\newtext{The following discussion on findings from this study is organized based on two application-specific tasks: the identification of important timesteps (T2) where bridge atoms exhibit distinct behavior and the analysis of their evolution patterns (T3), and the identification of patterns and timesteps (T2) associated with the formation of a new ring (T4).}

\myparagraph{Bridge atoms (T2, T3).}
\autoref{fig:CS2C1Evolution} shows the PC$_2$ 1D tracks for C1 and C8, annotated with the peeled CSPs at a few notable time steps. A distinct behavior is shown by the bridge atoms C1 and C8 around $48$~fs, highlighted in blue, which is not shown by any other atom corresponding to gray tracks. This is surprising because a major change in the electron density fields of one atom should impact the other atoms too. We analyze the evolution of bridge atoms and the bivariate field to understand the reason. The axes ($\phi_h, \phi_p$) of the annotated peeled CSPs are not shown to avoid clutter. The molecular CSPs are shown in the second row to aid in understanding the behavior of the bivariate field over the molecule. The donor strength~\cite{Sharma2023}, equal to $\phi_h^2 - \phi_p^2$, is computed and visualized using isocontours at $0.004$ and $-0.004$. The corresponding donor~(green) and acceptor~(red) fiber surfaces are shown in the bottom row. A negative donor strength implies a gain of charge. The green fiber surfaces highlight regions where $\phi_h > \phi_p$. The supplementary video shows the tracks and evolution of CSPs of the bridge atoms together with the fiber surfaces. 

We observe that the peeled CSP of C1 is diagonally aligned at $0$~fs. The molecular CSP also has a diagonal character. The green and red fiber surfaces are located near the bridge atoms and the atoms in the vicinity of the bridge atoms. It indicates that the bridge atoms are comparatively more active than those in the rings. This behavior is observed at all times when the molecular CSP is aligned diagonally, resulting in higher values of the moments for C1 and C8, making the tracks span a higher vertical range in the PCA 1D plot. Around $48$~fs, the tracks of C1 and C8 show a large vertical movement. The peeled CSP has a visible increase in size at those times and exhibits a vertical alignment. The molecular CSP also aligns with the axes. The ring R1 is covered with red fiber surfaces implying that atoms from C2 - C7 have vertically aligned CSPs. Green fiber surfaces enveloping atoms in R2 imply horizontal CSPs for C9-C14. The fiber surfaces around C1  are larger in this geometry than the atoms in R1, which explains the reason for a bigger peak in the track compared to other atoms. We observe this behavior in C8 as well. It behaves as a stronger donor and the peaks in its track are in the opposite direction to those of C1. An isolated study of the track might suggest that atoms other than the bridge atoms are inactive. However, the presence of fiber surfaces around these other atoms indicates that they are active as well, although to a lesser extent compared to the bridge atoms.

\myparagraph{A newly formed ring (T2, T4).} While traditionally regarded as a prototype \textit{cis}-to-\textit{trans} photoisomerization system (as enabled by the antibonding nature of the $\pi^*$-orbital across the bridge), a recent joint theory--experiment study suggests the prevalence of the cyclization pathway, producing 4a,4b-dihydrophenanthrene.\cite{karashima2023} This is facilitated by the in-phase nature of the facing lobes of the C3 and C10 orbitals in the particle NTO, enabling the formation of a new C3-C10 bond.

Around $228$~fs, C3 and C10 move close to each other and form a bond. The presence of fiber surfaces at $228$~fs in \autoref{fig:CS2C1Evolution} between C3 and C10 indicates the same. The creation of this bond causes a new ring to form containing C1, C2, C3, C8, C9 and C10. The molecule shows a unique character at this point. Alternate atoms behave as donor and acceptor, unlike other axis-aligned instances where the entire ring exhibits the same behavior. After $228$~fs, the bivariate field for the molecule seems to oscillate between two patterns. The molecular CSP switches to the diagonal orientation at $231$~fs. This diagonal orientation corresponds to the case where the bridge atoms are part of the new ring, and is different from the ones that appear before $228$~fs. Their donor/acceptor characters are not as strong as observed in the earlier case. The electron density is distributed over all the atoms in the new ring. The patterns in the track also indicate this behavior.  Further, two gray curves cover a larger vertical span when compared to the yellow and green tracks in the time period after $228$~fs. The two gray curves correspond to C4 and C11.  Additional observations regarding the opposing behavior of atoms from the two rings, including C4 and C11, are discussed in the supplementary material.

\section{Limitations}
The CSP and moment computation in the proposed method are one-time processes, but consume significant amount of time. The storage requirement increases with the number of instances. Additionally, the method produces tracks that may be complex and difficult to analyze when the number of atoms is large. After dimensionality reduction, the analysis becomes highly interactive, enabling users to easily filter out tracks that are either not relevant or cause clutter. However, this step does rely on the domain expertise of the user. The four moments that constitute the 4D representation of a CSP are well-suited for describing atomic CSPs, but may not be sufficient to describe generic CSPs. An additional limitation is that the strategy for selecting the principal components, as described in detail in the supplementary material, requires manual intervention to analyze the weights assigned to each moment by the individual principal components. \newtextF{Further efforts are required to develop our proposed pipeline into a ready to use tool that can be incorporated into the analysis workflow of a chemist.} We also note that each step of the visual analysis pipeline has an associated information loss. The segmentation introduces hard boundaries that may affect the bivariate field analysis within individual segments. Projecting moment vectors onto a plane introduces additional loss. A detailed discussion of information transformations at various stages of the pipeline is provided in Section 7 of the supplementary material.

\section{Conclusions}
This paper introduced a novel method to visualize time-varying bivariate fields using tracks in a PCA plot to summarize the shape and evolution of CSPs, \newtextF{and hence study electronic structure evolution}. This representation allows the identification of interesting time steps, regions, and pattern evolution. Molecular case studies demonstrated how this method can reveal critical time steps \newtextF{of interest} in the evolving electronic structure during nuclear dynamics induced by photoexcitation. The approach provides a compact representation of dynamics, facilitating the identification of significant electronic changes along nuclear trajectories.

\newtext{Analyzing the evolution of electronic structures poses significant challenges for chemists, especially when dealing with a large number of time steps and atoms in a molecule. The track-based representation supports, for the first time, an analysis of the evolution of electronic fields restricted to atoms and hence a comparison of the evolution sequence. Previous methods in the literature have not reported the ability to support such studies. The visual analysis pipeline utilizes fiber surfaces to validate the findings. Additionally, the method identifies critical time steps that correspond to significant changes in electronic structure and highlights atoms that are highly active or inactive. Both capabilities are highly valuable for a chemist. Simplifying the analysis of electronic structure evolution accelerates downstream higher-level analysis and enables efficient comparative study of molecules. Such analysis can potentially aid in the design of novel materials or medicines in the future.}

We plan to address some of the limitations of the method in future work to enable its use in other applications. \newtext{Addressing both storage and compute requirements will enhance the usability of the method. CSP resolution significantly impacts the storage requirements. Lower resolution CSPs may suffice to capture the coarse features of interest, such as size and orientation. \newtextF{GPU-based implementations may accelerate the computation of image moments. Both ideas require further investigation.} Currently, the selection and filtering of tracks depends crucially on the user. Simple automatic criteria based on coverage or area occupied may be designed to filter out the tracks. Exploring and identifying principal components also remains a manual and knowledge-intensive process. Automating the dimensionality reduction process is also a topic for future work. We plan to explore the use of additional moments or other approaches such as pre-trained deep networks to describe CSPs. However, a high-dimensional descriptor may result in additional challenges for the subsequent dimensionality reduction step.} The geometry of tracks is influenced by aggregate losses, potentially affecting analysis. Fine-grained analysis can mitigate some issues but requires manual investigation.



\section*{Acknowledgments}{%
	\newtext{This work is partially supported by an Indo-Swedish joint network project: DST/INT/SWD/VR/P-02/2019 and VR grant 2018-07085, MoE Govt. of India, a grant from SERB, Govt. of India (CRG/2021/005278), the Dr. Ram Kumar IISc Distinguished Visiting Chair Professorship in EECS, the SeRC (Swedish e-Science Research Center), Wallenberg Autonomous Systems and Software Program (WASP) funded by the Knut and Alice Wallenberg Foundation, and the Swedish Research Council~(VR) grants 2019-05487, 2022-02871, and 2023-04806. The computations were enabled by resources provided by the Swedish National Infrastructure for Computing (SNIC) at NSC partially funded by the VR grant agreement no. 2018-05973 and also by the super-computing resource Berzelius provided by the National Supercomputer Centre (NSC) at  Linköping University and the Knut and Alice Wallenberg Foundation.}
}

\bibliographystyle{IEEEtran}
\bibliography{references}

\newpage

\begin{IEEEbiography}[{\includegraphics[width=1in,height=1.25in,clip,keepaspectratio]{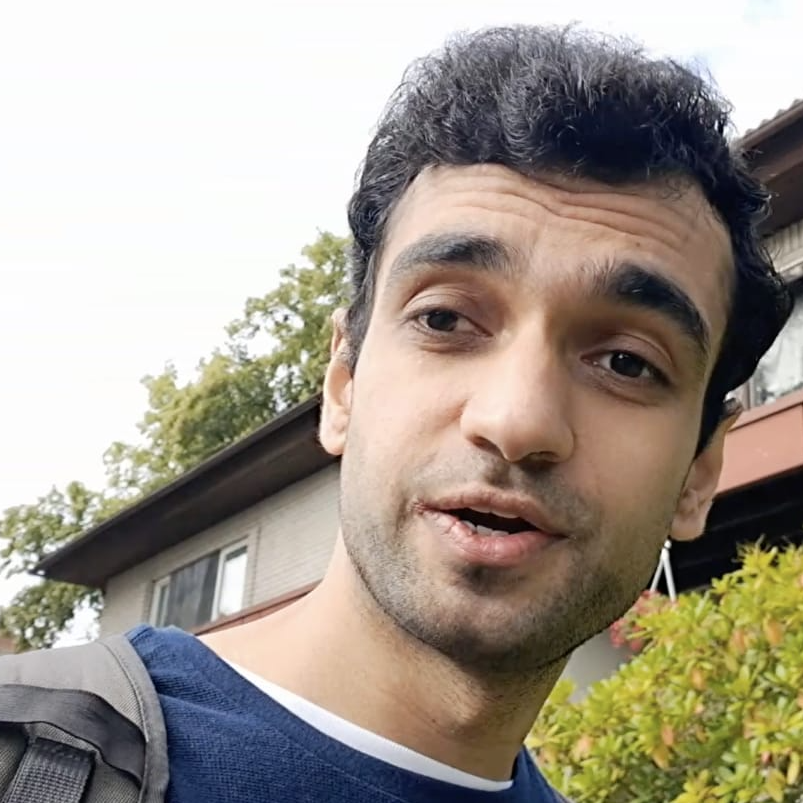}}]{Mohit Sharma}
is a postdoctoral fellow at Link\"{o}ping University, Sweden. He received his Ph.D. from the Computer Science and Automation Department at the Indian Institute of Science, Bangalore, and his M.Tech degree in Computer Science and Engineering from IIIT Hyderabad. His research interests include scientific visualization, computational topology, and its applications. Currently, he is working on multi-field data visualization.
\end{IEEEbiography}

\begin{IEEEbiography}[{\includegraphics[width=1in,height=1.25in,clip,keepaspectratio]{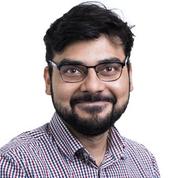}}]{Talha Bin Masood} is an assistant professor at Link\"{o}ping University in Sweden. 
He received his Ph.D. in Computer Science from the Indian Institute of Science, Bangalore.  His research interests include scientific visualization, computational geometry, computational topology, and their applications to various scientific domains.\end{IEEEbiography}

\begin{IEEEbiography}[{\includegraphics[width=1in,height=1.25in,clip,keepaspectratio]{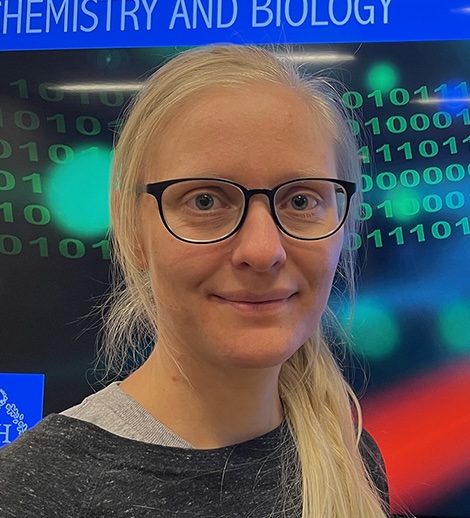}}]{Nanna Holmgaard List} is an assistant professor in theoretical chemistry at the Department of Chemistry, KTH Royal Institute of Technology. She received her Ph.D. degree from the University of Southern Denmark, Denmark. Her research revolves around theory development and application to understand light-induced processes in molecules.\end{IEEEbiography}

\begin{IEEEbiography}[{\includegraphics[width=1in,height=1.25in,clip,keepaspectratio]{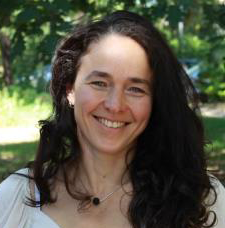}}]{Ingrid Hotz}
is currently a Professor in Scientific Visualization at the Link\"{o}ping University in Sweden. She received her Ph.D. degree from the Computer Science Department at the University of Kaiserslautern, Germany. Her research interests lie in data analysis  and  scientific  visualization,  ranging  from  basic  research questions  to  effective  solutions  to  visualization  problems  in  applications.  
\end{IEEEbiography}

\begin{IEEEbiography}[{\includegraphics[width=1in,height=1.25in,clip,keepaspectratio]{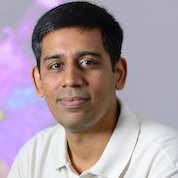}}]{Vijay Natarajan} is a Professor in the Department of Computer Science and Automation at Indian Institute of Science, Bangalore. He received the Ph.D. degree in computer science from Duke University. His research interests include scientific visualization, computational topology, and computational geometry. In current work, he is developing topological methods for time-varying and multi-field data visualization, and studying applications in biology, material science, and climate science.
\end{IEEEbiography}

\vfill

\end{document}